\documentclass{aa1}
\usepackage{graphicx}
\usepackage{color}
\usepackage{natbib}

\def\rsun{R_{\odot}}
\def\msun{M_{\odot}}

\begin{document}
\sloppypar
 
\title{Effect of the reverse shock on the parameters of the observed X-Ray
emission during the 1998 outburst of CI Cam}

\offprints{kate@iki.rssi.ru}

\author{E. V. Filippova\inst{1,2*}, M. G. Revnivtsev\inst{1,3}, 
and A. A. Lutovinov\inst{1}}

\institute{ 
    Space Research Institute, ul. Profsoyuznaya 84/32, Moscow, 117997 Russia
\and 
Max-Planck Institut f\"ur Astrophysik, Karl Schwarzschild Strasse 1, 
86740 Garching-bei-M\"unchen, Germany
\and
Excellence Cluster Universe, Technische Universit\"at M\"unchen, 
James-Franck-Strasse 1, D-85748 Garching,Germany
          }

        \authorrunning{Filippova et al.}
          \titlerunning{ }
\date{Received: 10 February 2009.}

\abstract{Based on the model of interaction between 
spherically symmetrical expanding matter and the
external medium, we have estimated the parameters 
of the matter heated by the shock that was produced
in the envelope ejected by the explosion of a 
classical nova during its interaction with the 
stellar wind from
the optical companion. Using this model, we have 
shown that the matter ejected during the outburst in
the system CI Cam had no steep velocity 
gradients and that the reverse shock could heat the
ejected matter only to a temperature of $\sim0.1$ keV. 
Therefore, this matter did not contribute to the mean
temperature and luminosity of the system observed 
in the energy range 3-20 keV.
\keywords{classical novae -- X-ray emission -- 
numerical simulations.} }

\maketitle

\section{Introduction}

Classical nova outbursts can be accompanied by
emission in both the standard (1 - 10 keV) and hard
($>$ 20 keV) X-ray energy bands. Matter heated by
the shock that was produced by a high-velocity 
($\sim 1000-4000$ km/s) expansion of the envelope of a
white dwarf is believed to be the source of this X-ray
emission. In the system CI Camelopardalis (CI Cam),
the optical companion is a B[e]-type B4III-V star
(Barsukova et al. 2006) with a strong stellar wind
(Robinson et al. 2002; Filippova et al. 2008), which
produces a dense circumstellar medium around the
white dwarf. Filippova et al. (2008) (hereafter Paper I)
showed that, in this case, a shock is generated by
the envelope expansion in the circumstellar medium,
which could heat a large amount of 
stellar wind matter up to 10-20 keV 
sufficient to produce a high X-ray luminosity.

In the envelope itself, a shock (or initially a reverse
rarefaction wave that will transform into a reverse
shock as the envelope expands) will also form.
Under certain conditions, this shock can also heat the
matter to high temperatures. For example, it follows
from the analytical calculations by Chevalier (1982)
and Nadyozhin (1985) that when the ejected matter
interacts with a constant-density medium, the
temperature at the reverse shock for an
power in the ejected matter density profile of 
$\sim 6 - 8$ can be lower than the temperature at the forward
shock only by a factor of $\sim 2 - 4$.

A schematic view of the system of shocks produced
by the interaction of ejected matter with the
circumstellar medium is shown in Fig. 1 (the outer
boundary of the envelope is a contact discontinuity).
In general, the formation time and the law of motion
of the reverse shock depend on the matter velocity and
pressure distributions inside the envelope. However,
at present, there is no complete model that would
consistently describe the evolution of the profiles of
these parameters in the matter ejected by the explosion
of a classical nova (see, e.g., a review of theoretical
models in Friedjung (2002)). On the basis
of numerical calculations and their comparison with
observational data during a classical nova explosion,
two mechanisms of matter ejection at the initial time
are suggested: under the action of thermal pressure
and through a shock wave. The consequences of the
ejection of matter through these mechanisms were
considered by Sparks (1969). He showed that in the
case of pressure-driven expansion, the envelope has
a very shallow matter velocity gradient, while in the
case of shock-driven ejection of matter, it expands
with a steep velocity gradient. At later expansion
stages, after the maximum optical brightness  of the
nova, the expansion of matter is described by the
model of an optically thick wind (Kato and Hachisu
1994; Hauschildt et al. 1994). There also exist theories
predicting that shortly after the maximum optical 
brightness, the velocities of the envelope layers closer
to the white dwarf are higher than those of the outer
layers (McLaughlin et al. 1947, 1964).

An example of the possible development of a classical
nova explosion was given by Prialnik (1986),
who calculated a complete cycle of the evolution of a
classical nova explosion, from the phase of accretion
to its resumption; the conditions for the generation
of a shock wave in the envelope were met. According
to these calculations, within the first half an hour 
($\sim 2000$ s) after the thermonuclear explosion, the white
dwarf photosphere expands to $\sim 10 \rsun$ due to the
shock breakout. Within the next $\sim 4$ h, the envelope
expands even more due to the radiation pressure from
the white dwarf surface, with the optical flux from the
system reaching its maximum. It follows from observations
that the characteristic time it takes for the
optical flux to reach its maximum for most classical
novae is $< 3$ days, but exceptions are also observed;
for example, during the outburst of Nova LMC 1991
this time was $\geq 13$ days (Schwarz et al. (2001) and
references therein). According to calculations, the
outer layers of the envelope at this time expand with
constant velocities, which increase toward the outer
boundary and reach $\sim 3800$ km/s at it. This part of
the envelope ceases to be connected with the white
dwarf and expands by inertia, interacting only with
the circumstellar medium. The remaining part of the
envelope initially contracts under the gravitational
force and, after some time, again begins to expand
under radiation pressure in the regime of an optically
thick wind.

It was shown in Paper I that the main peculiarities
of the behavior of the light curve and radiation
temperature during the X-ray outburst of CI Cam in
the first (spherically symmetric) approximation could
be described in terms of the radiation model of stellar
wind matter heated by the forward shock produced in
a classical nova explosion. In this model, the envelope
is ejected from the white dwarf due to explosive
thermonuclear burning, which already on 0.1-0.5 day
after the explosion onset has an expansion velocity
of $\sim 2700$ km/s and flies under the action of an
external force, for example, the radiation pressure
from the white dwarf, with a constant velocity for the
first $\sim 1-1.5$ days. Subsequently, the envelope probably
becomes transparent and decelerates, interacting
with the matter of the stellar wind from the optical
companion. Based on a comparison of the observed
rise in luminosity with the theoretical dependence,
we estimated the stellar wind density near the white
dwarf to be 
$n_0(r \le r_c)\sim 8.6\times 10^9d_{2kpc}U^{-3/2}_{2700} cm^{-3}$,
which transforms into the law $n_0 \sim r^{-2}$ at 
$r > r_c = 1.9\times10^{13}$ cm. 
In the simplest model, this stellar
wind density distribution corresponds to a mass loss
rate of the optical star $\sim (1-2)\times 10^{-6} \msun/yr$. The
observed time dependence of the temperature of the
emitting matter at late envelope expansion stages allowed
us to constrain the mass of the ejected envelope
based on our model, $10^{-7}-10^{-6}\msun$. Note that in this
model, the processes in the envelope itself were disregarded;
in our calculations, we used a finite-mass
piston as the envelope.

In this paper, we calculated the contribution from
the emission of the ejected envelope matter heated
by the reverse shock to the observed radiation temperature
and luminosity of CI Cam during its X-ray
outburst in 1998.

\begin{figure}[htb]
\centerline{\includegraphics[width=0.8\columnwidth,bb=0 -5 140 160,clip]{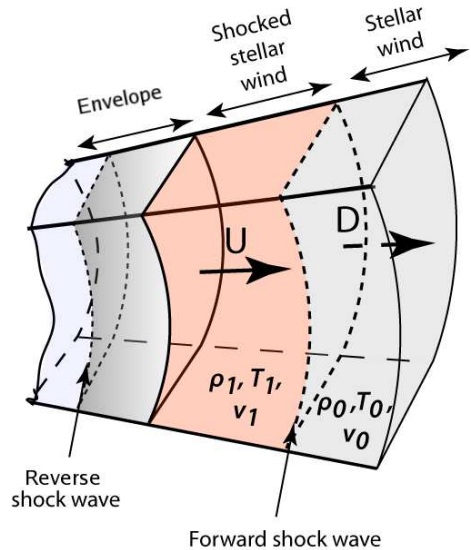}}
\caption{Scheme of the shocks produced by the interaction
of the expanding envelope with the circumstellar medium:
$U$ is the velocity of the contact discontinuity or the envelope
(depending on the model considered) and $D$ is the
velocity of the forward shock.}\label{scheme}
\end{figure}

\section{Numerical calculations}

To model the reverse shock, we used a numerical
scheme described in Paper I: a one-dimensional,
spherically symmetric code in Lagrangian coordinates
with a staggered mesh (the cell radius, velocity,
and mass are determined at the cell boundaries,
while the density, pressure, and internal energy are
determined at the cell centers).

At the initial time, the outer boundary of the envelope
was placed at a distance of $10^{12}$ cm. The density
of the circumstellar medium was specified as follows:
$n_0 = 8\times10^9 cm^{-3}$ at $r < r_c$ and $n_0\sim r^{-2}$ 
at $r > r_c$. The initial cell size 
was $\Delta r = 10^{10}$ cm. The velocity
of the matter at the inner and outer boundaries was
specified by a time-independent constant.

As in Paper I, we took into account the radiative
cooling of the matter heated by the shocks in an
optically thin regime.

As was shown in Paper I, the matter behind 
the shocks is a multitemperature plasma in the
sense that the plasma temperature is nonuniform
along the radius. Consequently, the radiation temperature
that we measure based on X-ray observations
is an average quantity and it may not be equal
to the temperature at the shock front. Therefore, to
obtain the calculated mean temperature, we used the
same averaging procedure as that for observations.
We calculated the ratio of the fluxes in the 3-5 
and 5-20 keV energy bands, which, in turn, 
corresponds to a
certain temperature in the radiation model 
of a single temperature,
optically thin plasma. The method is
described in more detail in Paper I.

\begin{figure*}[htb]
\centerline{\hbox{
\includegraphics[width=0.95\columnwidth,bb=20 145 580 510,angle=0,clip]{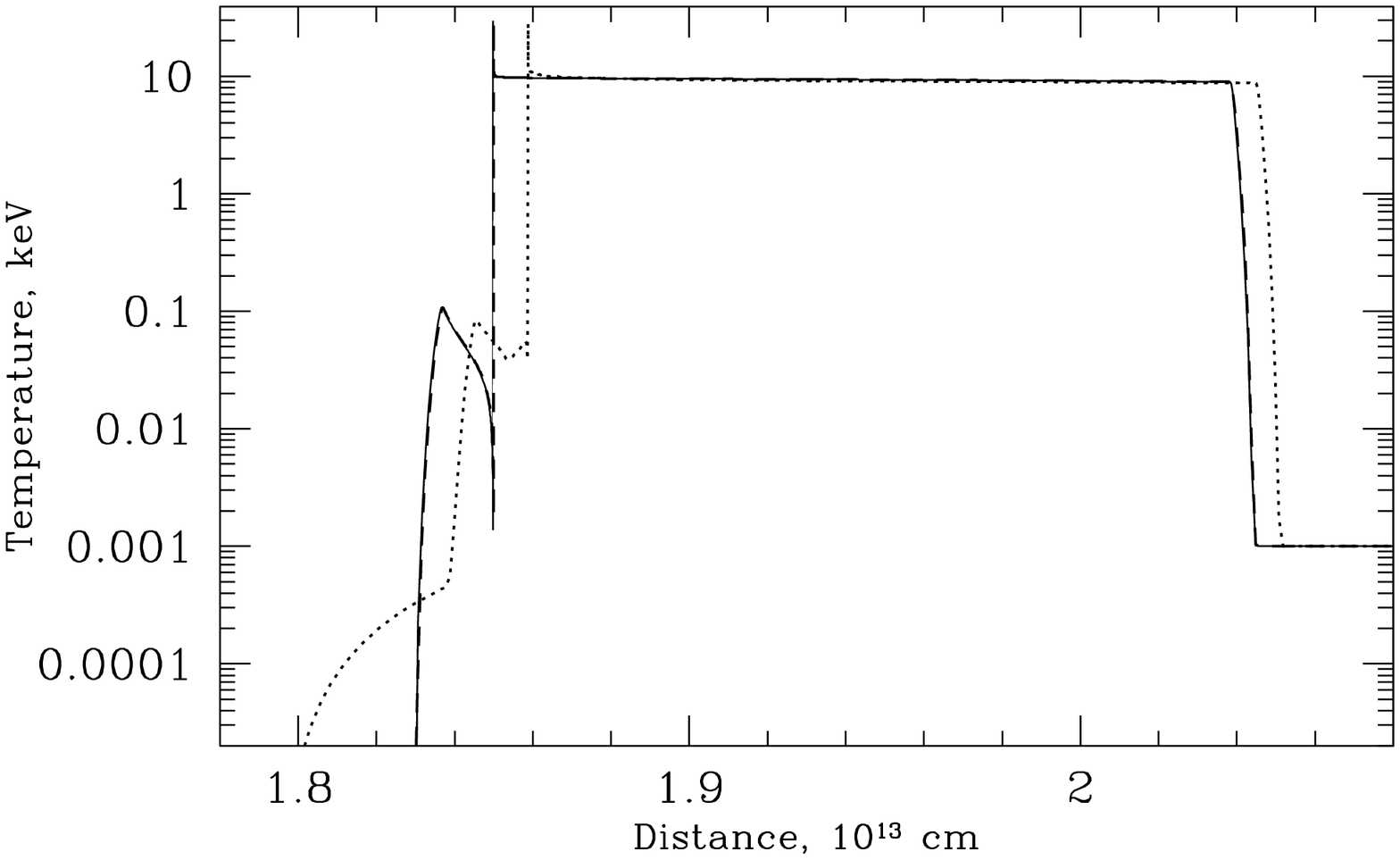}
\includegraphics[width=0.95\columnwidth,bb=20 145 575 510,angle=0,clip]{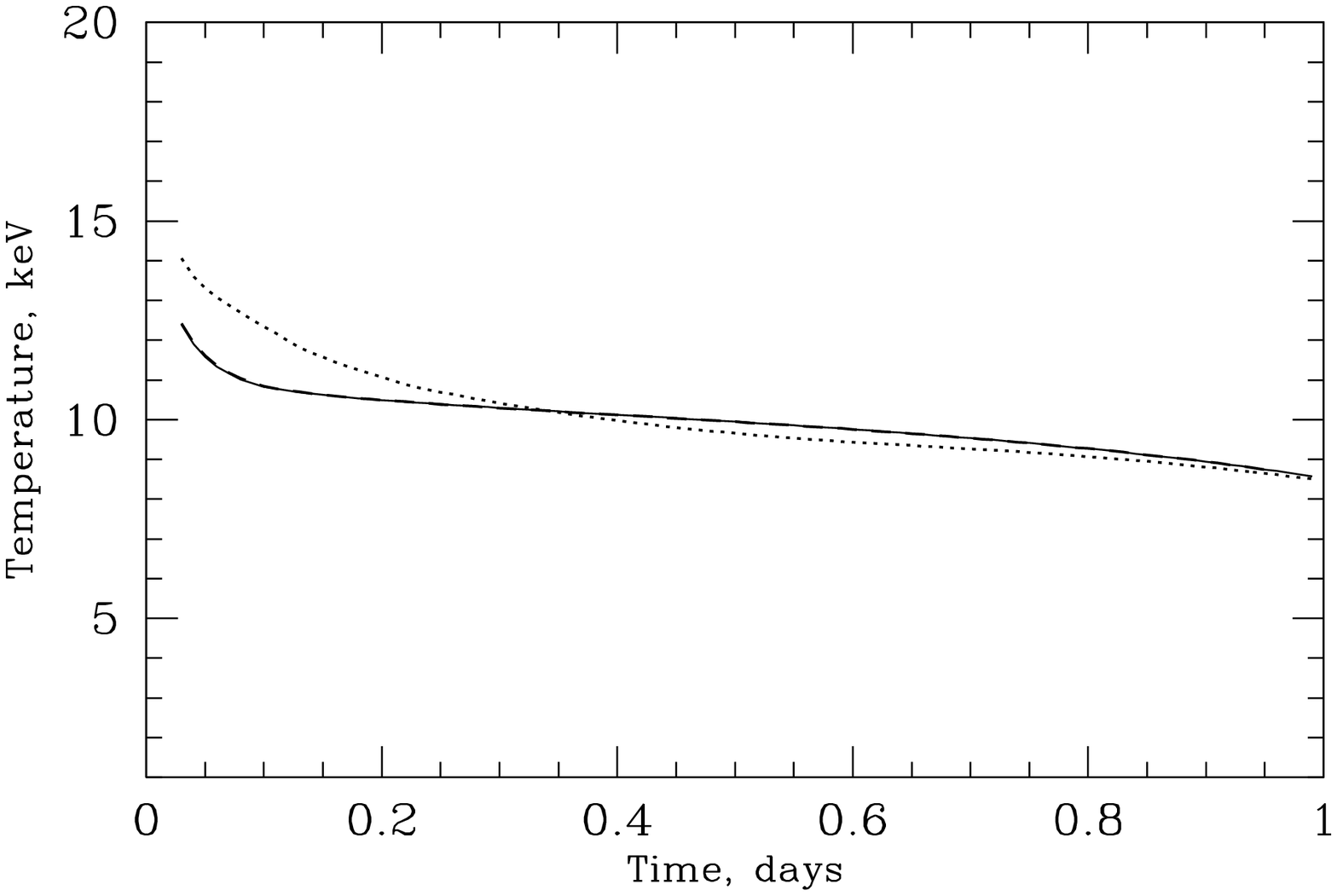}
}}\caption{Left panel: temperature profiles in the envelope and 
stellar wind on 0.8 day after the explosion for various initial envelope
matter temperatures; the solid, dashed, and dotted lines 
correspond to $10^3, 10^4$ (in both cases, the contact discontinuity is at
a distance of $\sim 1.85\times10^{13}$ cm), 
and $10^6$ K (the contact discontinuity is at a distance 
of $\sim 1.86\times 10^{13}$ cm), respectively; right panel:
time dependence of the mean temperature for these cases.}\label{tob}
\end{figure*}

\subsection{Effect of the matter temperature in the envelope
on the calculated parameters}

The temperature or pressure of the envelope matter
is a parameter that, in general, can affect the
formation and propagation of shocks after the decay
of an arbitrary discontinuity.

To understand what the temperature and pressure
distributions in the ejected envelope are, we can turn
to actual observations of classical novae. Two outbursts
of novae (Cyg 1992 and LMC 1991) that were
observed before the maximum optical brightness was
reached and for which the radiation temperature was
measured are known to date. However, the effective
radiation temperature obtained in such an analysis of
observations is not a good indicator of the physical
temperature in the envelope (Hauschildt et al. 1994;
Schwarz et al. 2001). Nevertheless, since the emission
from the envelope matter has a maximum in the
ultraviolet, we may assert that the matter temperature
does not exceed $\sim 0.1$ keV in order of magnitude.

To answer the question of how the envelope matter
temperature affects the propagation of shock waves,
we performed calculations with the following initial
parameters of the matter in the envelope: the density
is constant along the radius (the envelope mass
was $10^{-6}\msun$, the velocity is also constant along the
radius and equal to $2700$ km/s, and we considered
several matter temperatures: $10^3, 10^4$, and $10^6$ K.

Figure 2 (left panel) shows the matter temperature profiles
in the envelope and stellar wind on 0.8 day after the
onset of expansion: the dotted, dashed, and solid lines
correspond to $T = 10^6, 10^4$, and $10^3$ K, respectively.
It follows from the figure that the expected range
of envelope matter temperatures affects weakly the
propagation dynamics and strength of the forward
shock and leads to unimportant differences in the
time dependence of the mean radiation temperature
(Fig. 2 (right panel)). Therefore, below, the initial temperature of
the envelope matter in our calculations was set equal
to a constant, $10^4$ K, unless stated otherwise.

\section{Transformation of the reverse rarefaction wave into
a reverse shock}

As was noted in Paper I, the interaction of the
ejected envelope with the circumstellar medium can
give rise to a reverse rarefaction wave at the very
outset. The instant of the subsequent transformation
of the reverse rarefaction wave into a reverse shock
depends on the radial distribution of envelope matter
parameters (such as the velocity and pressure). In the
same paper, we made very simple estimates of the
conditions under which the reverse rarefaction wave
is generated and the time when it transforms into a
reverse shock. It follows from these estimates that
the reverse rarefaction wave transforms into a shock
almost immediately.

For a clear demonstration of this phenomenon, we
performed calculations in which this transformation
could be traced in more detail. It should be noted
that the parameters specified as the initial conditions
bear no relation to the actual values: for example, in
order that the reverse rarefaction wave could recede
noticeably from the contact discontinuity, the matter
temperature at the outer boundary of the envelope
was set equal to 10 keV, but in order that the disturbances
arising at the inner boundary have no time
during the calculations to propagate over the entire
envelope, the matter temperature was specified by a
linear function of the radius and was $\sim 0.01$ keV at
the inner boundary; the energy losses through radiation
were disregarded. The expansion velocity of the
envelope matter was 600 km/s.

Figure 3 shows the temperature and pressure profiles
obtained in this model in the interacting region
at various times. We clearly see how the rarefaction
wave is generated (solid line) and how it transforms
into a reverse shock (in the profiles drawn by the
dashed line, the reverse shock is seen clearly). The
cell number is along the X axis, with the contact
boundary being located on cell no. 800; the forward
shock propagates through the stellar wind rightward;
the rarefaction wave propagates through the ejected
matter leftward and transforms into a shock.

\begin{figure*}[htb]
\centerline{\hbox{
\includegraphics[width=0.95\columnwidth,bb=20 145 595 510,angle=0,clip]{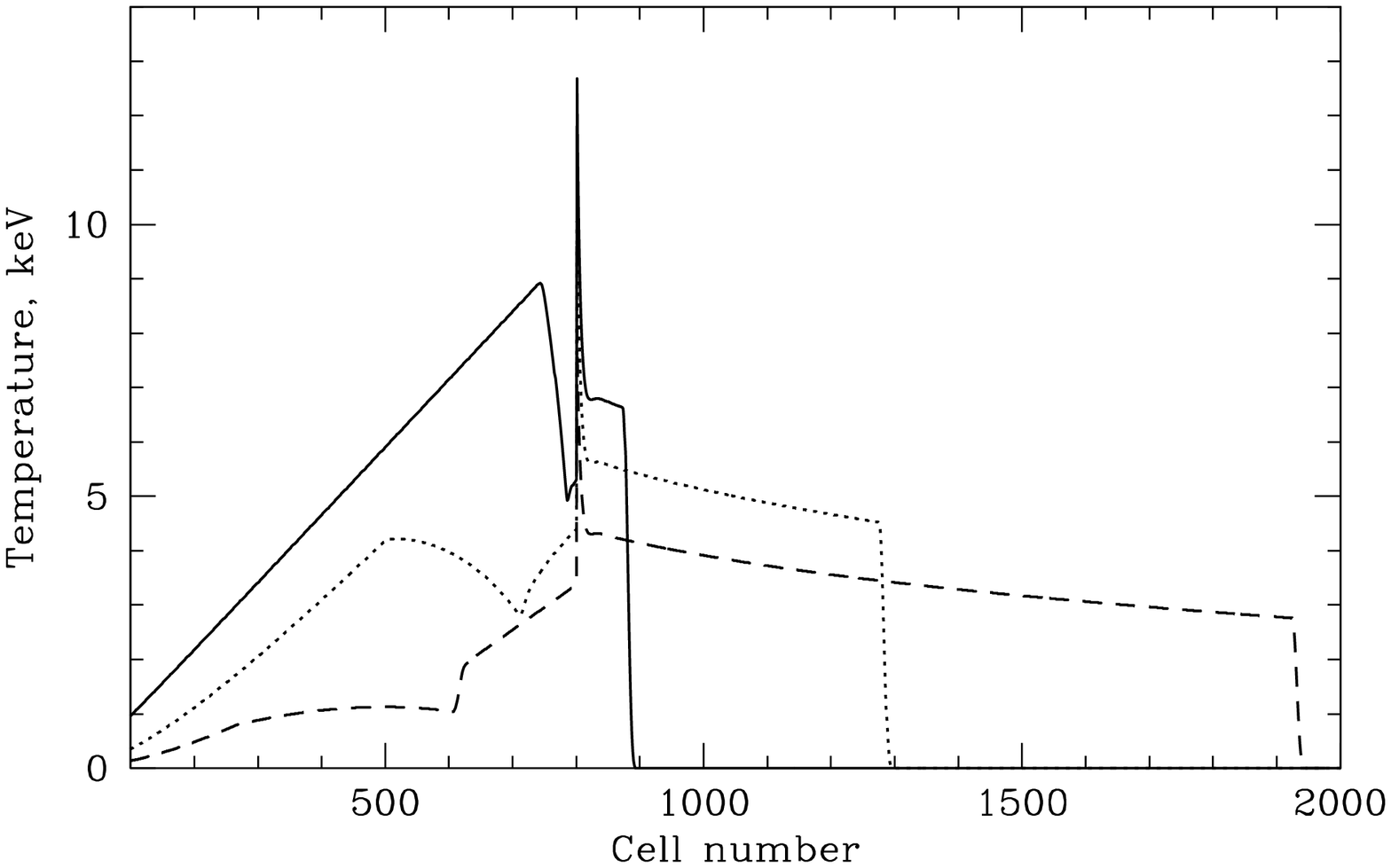}
\includegraphics[width=0.95\columnwidth,bb=20 145 595 510,angle=0,clip]{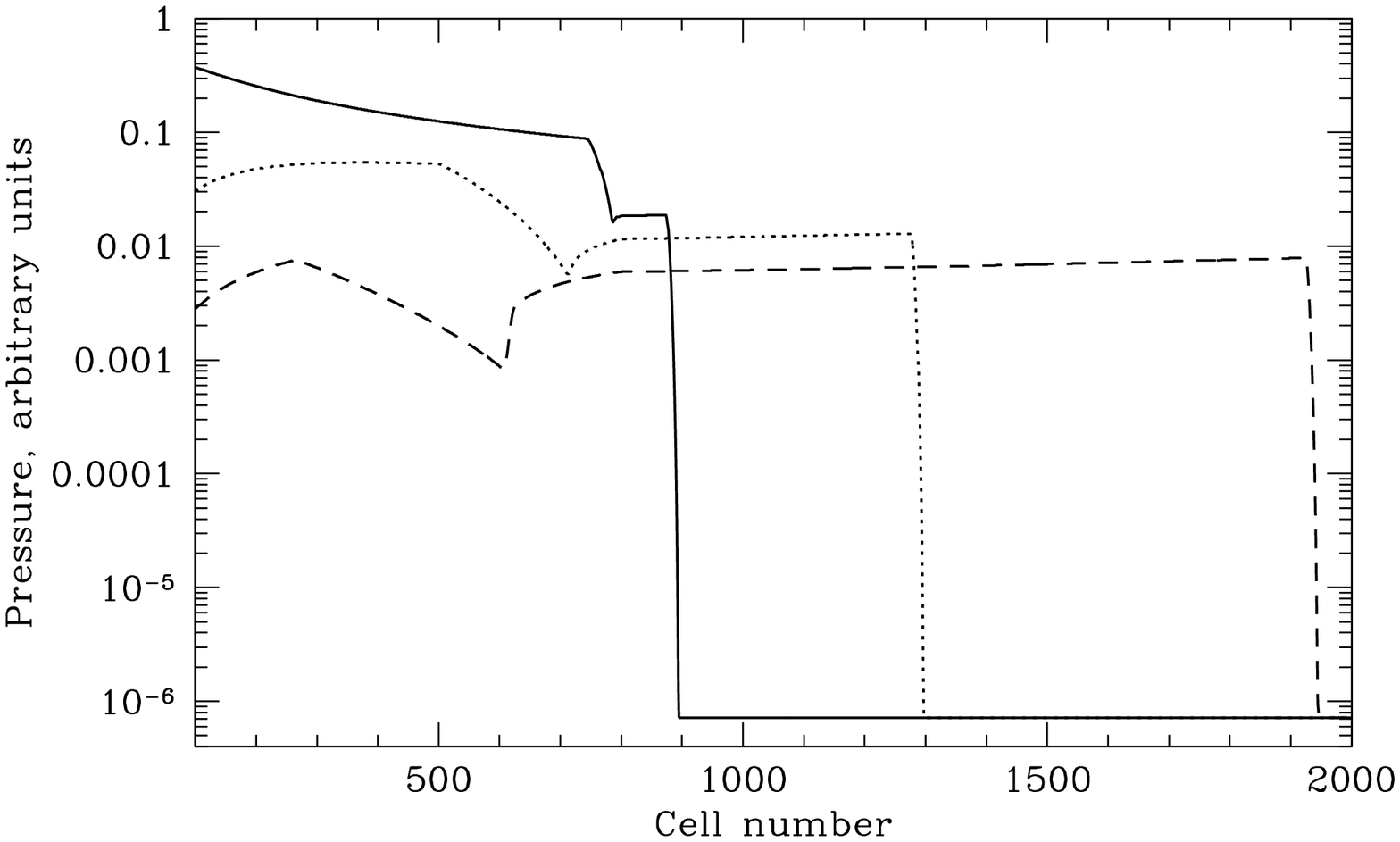}
}}
\caption{Left panel: temperature profiles in the ejected 
matter and stellar wind at various times; the cell number 
is along the X axis, the
contact boundary is located on cell no. 800. Right panel: 
pressure profiles at the same times.}\label{rev_shock}
\end{figure*}

\section{Homologous envelope expansion $v \sim r$}

Classical nova explosions resemble in mechanism
type Ia supernova (SN Ia) explosions, thermonuclear
explosions of white dwarfs (Woosley and Weaver
1986). However, in the former case, the explosion
energy and, hence, the kinetic energy of the ejected
envelope are much lower than those in the latter case.
For classical nova explosions and supernovae, the
kinetic energy of the ejected envelope is estimated to
be $\sim 10^{44}-10^{45}$ erg (Starrfield et al. 1976) 
and $\sim 10^{51}$ erg (Khokhlov et al. 1993), respectively.

Numerical calculations show that a homologous
expansion of the ejected matter during an SN Ia
explosion is established in $\sim 10$ s (R\"opke et al. (2005)
and references therein). Dwarkadas et al. (1998) provided
the density profiles of the ejected matter for
several SN Ia explosion models used to describe the
observational data. They also showed that the density
profiles could be described both by a power law with
an exponent of 7 (nevertheless, the exponent can often
be different from 7) and by an exponential law 
$\rho \sim e^{-v/v_0}$.

Chevalier (1982) and Nadyozhin (1985) provided
self-similar solutions for the decay of an arbitrary
discontinuity in the case of a spherically symmetric
homologous envelope expansion with powers in
the density distribution $p > 5$ into a constant-density
external medium. It follows from these solutions that
when the reverse shock is produced, the temperature
at the forward shock decreases with time. For example,
for $p=5.4$, the time dependence of the temperature
at the forward shock is $T_1\sim(t/t_{min})^{-10/9}$
(Nadyozhin 1985). For an envelope with a mass of
$10^{-6}\msun$, its kinetic energy is $10^{44}$ erg 
and the density
of the matter of the external medium is $8\times10^9$ cm$^{-3}$,
the time $t_{min}$ is $\sim 9000$ s; we then find 
from the formula
that in the first day of expansion, the temperature
at the forward shock will fall by a factor of $\sim 10$,
which will lead to a decrease in the mean radiation
temperature.

To understand how the homologous envelope expansion
in our problem affects the behavior of the
mean radiation temperature, we performed the following
numerical calculations. First, we considered
the case where the mass loss rate from the white
dwarf as a result of the shock breakout was constant.
The power in the radial density distribution of the
ejected matter is then p = 3. Since the velocity of
the ejected matter decreases with decreasing radius,
only the high-velocity outer layers of the envelope are
actually of interest for the generation of a forward
shock and energy estimations for the Sedov phase.
For our subsequent estimations of the kinetic energy
and mass of the envelope, we took a lower velocity
limit of 2000 km/s.

We performed our calculations for two models:
the mass of the matter ejected with a velocity
$>2000$ km/s is $\sim 10^{-6}\msun$ and 
$\sim 2\times10^{-5}\msun$.
The kinetic energy of the envelope is 
$E_{kin,v > 2000}\sim 7\times10^{43}$ erg in 
the former case and $E_{kin,v>2000}\sim1.2\times10^{45}$ erg 
in the latter case. In calculating the
mass, we used the cosmic abundance of the matter.
However, since the heavy-element abundance in the
ejected envelope is believed to be higher than the
cosmic one (Starrfield et al. 1976; Yaron et al. 2005),
the obtained mass and energy are lower limits.

Figure 4 shows the time dependence of the mean
temperature of the matter behind the forward
shock for these two cases (the solid and dashed lines
correspond to masses of the matter with a velocity
$>$ 2000 km/s of $2\times10^{-5} \msun$ and $10^{-6}\msun$, 
respectively). In Paper I, we estimated the explosion energy
required to obtain the observed time dependence of
the mean temperature of the matter at late expansion
stages, when the shock enters Sedov regime
(the regime in which the shock ``forgets'' the details of
its formation and evolves self-similarly). 
According to these calculations, the energy 
should be $\sim (5 - 10)\times 10^{43}$ erg. 
Thus, a further increase in the mass
of the ejected matter in the model with a homologous
expansion will lead to a higher (compared to the
observations) radiation temperature on 4 - 10 days of
expansion.

In Hauschildt et al. (1994) and Schwarz et al.
(2001), who investigated the structure of the envelopes
ejected by the explosions of the classical novae
Cyg 1992 and LMC 1991 in the initial expansion
period, the constructed models that described best
the data, had power-law density profiles with powers
of 15 and 7, respectively. In our case, a steeper
density distribution in the outer parts of the ejected
envelope only compounds the situation, because the
problem has a constraint on the energy of the outer
layers of the envelope. The velocity of the matter at
the outer boundary should remain constant, because
it determines the temperature at the forward shock.
Therefore, when the parameter $p$ is varied, the mass
of the outer layers with a velocity $> 2000$ km/s
should be retained. When $p$ increases, this will lead
to a decrease in the density at the outer boundary of
the ejected matter at the initial time and, hence, the
reverse shock will develop more intensively for some
time. The dotted line in Fig. 4 indicates the time dependence
of the mean temperature for the case where
the mass of the matter with a velocity $> 2000$ km/s
is $2\times10^{-5}\msun$ and p = 15.

We see from the figure and the above estimates
that the case with a homologous expansion of the
envelope matter in our one-dimensional model is in
conflict with the observational data.

The absence of a homologous expansion phase
during the outburst in the system CI Cam can be
explained in several ways. The conditions for the generation
of a shock wave in the envelope may have not
been met during the explosion; therefore, the matter
was ejected by thermal pressure without any velocity
gradient (Sparks 1969). Or it is possible that the
matter ejected as a result of the shock breakout was
almost immediately stopped due to the high density
of the external medium. The shock produced by it
was damped out almost immediately and the observed
shock was generated by the subsequent ejection of
matter due to the radiation pressure.

\begin{figure}[htb]
\includegraphics[width=0.95\columnwidth,bb=20 145 575 510,angle=0,clip]{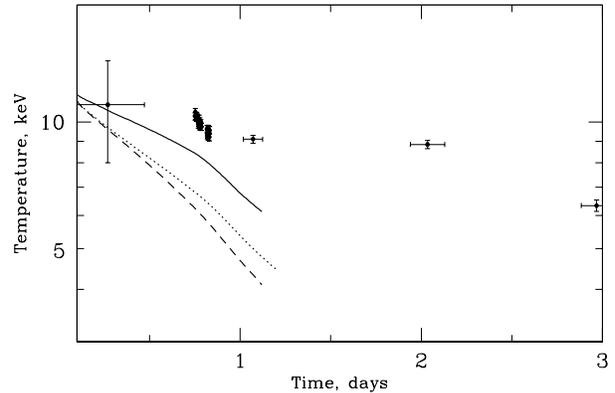}
\caption{Time dependence of the mean temperature of the matter 
behind the forward shock for a homologous expansion
of the matter in the envelope: the solid line correspond to 
the model with a mass of the matter in the envelope having a velocity
$> 2000$ km/s of $2\times10^{-5}\msun$ and p = 3; the dashed line 
corresponds to $10^{-6}\msun$, the exponent in the density distribution
of the ejected matter is p = 3; the dotted line corresponds 
to $2\times10^{-5}\msun$, p = 15.}\label{vr}
\end{figure}

\section{Envelope expansion with a constant velocity 
of matter}

We performed calculations for three density profiles
of the matter in the envelope at the initial time:
$\rho = const, \rho \sim r^{-2}$, and $\rho \sim r^{-3}$. 
The envelope mass
was taken to be $\sim 10^{-6}\msun$. 

To keep the velocity of the forward shock during
the decay of an arbitrary discontinuity the same as
that when the piston envelope is pushed, the velocity
of the matter in the envelope should be higher than the
piston velocity. In these calculations, 
we set it equal to 3000 km/s.

The derived temperature profiles of the matter
behind the forward and reverse shocks are
shown in Fig. 5. It follows from this figure that in
all cases the temperature behind the reverse
shock does not exceed 0.1 keV (recall that radiative
cooling in our model switches on at temperatures
$> 0.1$ keV), i.e., this matter does not contribute to
the observed flux in the 3-20 keV energy band and to
the temperature averaged over the X-ray flux during
approximately the first 0.7 day of expansion.

Investigation of the behavior of the reverse shock
at later times is hindered by the absence of reliable
theoretical models for the distribution 
of physical parameters in the expanding matter.

Note that the envelope expanded freely in these
calculations. However, as was shown in Paper I, on
the first day of expansion, an external force keeping
its velocity constant should act on the matter in the
envelope. It is quite possible that the reverse shock
will be suppressed even more in this case.

We see from the figure that the radiative cooling
of the matter behind the reverse shock is
important only for the model with a density 
profile $\rho \sim r^{-3}$ (dashed line) - the temperature 
profile exhibits a
"shelf" behind the reverse shock. Let us show
that the radiative cooling in this case also takes place
in an optically thin regime.

\begin{figure}[htb]
\includegraphics[width=\columnwidth,bb=20 145 585 510,angle=0,clip]{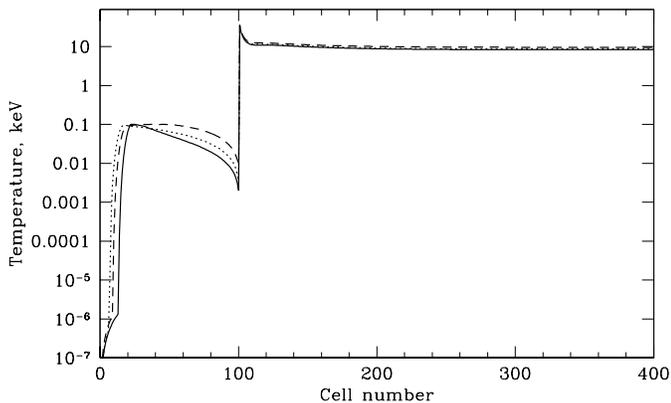}
\caption{Temperature profiles behind the forward 
and reverse shocks on 0.5 - 0.7 day after the explosion onset for an
envelope with a mass of $\sim 10^{-6}\msun$: the solid, dotted, 
and dashed lines correspond to $\rho = const, \rho \sim r^{-2}$, 
and $\rho \sim r^{-3}$,
respectively. The contact boundary is on cell no. 100.}\label{vconst}
\end{figure}

\subsection{Radiative cooling of the matter behind the reverse shock}

To establish the regime of radiation of the matter
behind the reverse shock, we calculated its
optical depth. The optical depth for Thomson scattering
up to the region behind the reverse
shock in which the radiative cooling becomes important
is less than unity. A significant contribution
to the absorption of radiation in the envelope matter
can be made by absorption in lines. However, using
the opacity tables calculated with the OPAL code, we
found that for a density of the order of 
$3\times 10^{12}$ cm$^{-3}$
and a temperature of $5\times 10^3 -  10^5$ K, the absorption 
cross section in the matter doesn't exceed 
$\sim 10^{-24}$ cm$^2$. The optical depth corresponding to this 
cross section is also less than unity. Thus, the matter
behind the reverse shock, just as behind 
the forward shock, radiates in an optically thin
regime.

As we have already said above, the heavy-element
abundance in the envelope matter ejected by the explosion
of a classical nova can be higher than the
solar one, which, in turn, can lead to an increase in
the cooling rate of the matter. In our calculations,
we retained the solar abundance, because the characteristic
cooling time for the density obtained in
the calculations ($3\times 10^{12}$ cm$^{-3}$) is 
$\tau_{rad}\sim 4$ s even
for the solar abundance. If it is lower by a factor
of several, then this will not affect the calculations.
It follows from the formulas in Paper I that for the
matter behind the reverse shock, the characteristic
time it takes for a Maxwellian velocity distribution
of ions at $T_i = 0.1$ keV to be established
is $\tau_{ii}\sim6\times10^{-4}$ s 
(in these estimates, we took the
parameters $A_i = 1, A_e = 1/1836, ln\Lambda= 15, Z_i = 1$,
and $n_i = n_e$, as in Paper I). Formally, for the ion and
electron temperatures $T_i = 0.1$ keV and $T_e = 1$ eV,
the ion Maxwellization time is longer than the time
of temperature equalization between the ions and
electrons by a factor of 
$\tau_{ii}/\tau_{ie} = 4.5\times10^{-2}(T_e/T_i+5\times10^{-4})^{-3/2}\sim42$. 
This means that the ions initially
transfer their energy to the electrons without
still having a Maxwellian velocity distribution, while
Spitzer's formula for $\tau_{ie}$ begins to work only when
$T_e > 0.13T_i \sim 0.01$ keV. For our estimations, we take
$T_e = 0.05$ keV (in this case, 
$\tau_{ee}\sim5\times10^{-6}$ s), then
$\tau_{ie}\sim5\times10^{-3}$ s. Clearly, 
the time of electron heating
by ions to $T_e = 0.05$ keV should be of the order of
the value of $\tau_{ie}$ obtained. The time it takes for an
ionization equilibrium to be established is $\tau_{eq} \sim 1 s$.
Thus, it follows from our estimates that the characteristic
times it take for an equilibrium to be established
behind the reverse shock are shorter
than the characteristic radiative cooling time, $\sim 4$ s,
and the applicability conditions for the APEC model
(http://hea-www.harvard.edu/APEC/REF) to calculate
the plasma energy loss rate are met\footnote{The same 
reasoning should also applied to the estimates
made in Paper I.}.

\section{Conclusions}

We investigated the effect of the reverse shock on
the observed parameters of the X-ray emission during
the 1998 outburst of CI Cam using a spherically
symmetric model for the interaction of the envelope
ejected by the nova explosion with the circumstellar
matter. Comparison of our numerical calculations
and observations in the frame of this model led us to the
following conclusions.

\begin{itemize}

\item Te homologous expansion phase of the matter
during the explosion in CI Cam most likely was either
absent or short and did not give rise to an observable
forward shock in the stellar wind. The velocity profile
in the matter ejected by the explosion had no steep
gradients.

\item For a free envelope expansion at a constant
velocity and the explosion parameters that we obtained
in Paper I, the reverse shock could not heat the
matter to temperatures above $\sim 0.1$ keV during the
first $\sim0.7$ day of expansion.

\item During the 1998 outburst of CI Cam, the contribution
from the matter heated by the reverse shock
to the observed luminosity in the 3 - 20 keV energy
band and to the temperature averaged over the X-ray
flux during the first $\sim0.7$ day of envelope expansion
was negligible compared to the contribution from the
matter behind the forward shock.
\end{itemize}

\subsection*{Acknowledgments}
This work was supported by the Russian Foundation
for Basic Research (project no. 07-02-01051),
the Program of the Russian President for Support
of Scientific Schools (NSh-5579.2008.2), and the
``Origin, Structure, and Evolution of Objects of the
Universe'' Program of the Presidium of the Russian
Academy of Sciences. E.V. Filippova is also grateful
to the Foundation for Support of Russian Science
for support. We wish to thank the referees for their
useful and important remarks, which helped to improve
significantly the paper, and S.A. Grebenev for
a discussion of the obtained results.

\section*{References}

1. E. A. Barsukova, N. V. Borisov, A. N. Burenkov, et al.,
Astron.Rep. 50, 664 (2006).

2. R. Chevalier, Astrophys. J. 258, 790 (1982).

3. V. Dwarkadas and R. Chevalier, Astron. Astrophys.
497, 807 (1998).

4. E. Filippova, M. Revnivtsev, and A. Lutovinov, Astron.
Lett. 34, 797 (2008).

5. M. Friedjung, AIP Conf. Ser. 266 (2002).

6. P. Hauschildt, S. Starrfield, S. Austin, et al., Astrophys.
J. 442, 831 (1994).

7. M. Kato and I. Hachisu, Astrophys. J. 437, 802
(1994).

8. F. Khokhlov, E.M\"uller, and P. H\"oflich, Astron. Astrophys.
270, 223 (1993).

9. D. McLaughlin, Publ. Astron. Soc. Pacific 59, 244
(1947).

10. D.McLaughlin, Ann. d'Astrophys. 27, 496 (1964).

11. D. Nadyozhin, Astrophys. Space Sci. 112, 225
(1985).

12. D. Prialnik, Astrophys. J. 310, 222 (1986).

13. E. L. Robinson, I. I. Ivans, and W. F. Welsh, Astrophys.
J. 565, 1169 (2002).

14. F. K. R\"opke, Astron. 432, 969 (2005).

15. G. Schwarz, S. Shore, S. Starrfield, et al., Mon. Not.
R. Astron. Soc. 320, 103 (2001).

16. W.M. Sparks, Astrophys. J. 156, 569 (1969).

17. S. Starrfield, W. M. Sparks, and J. W. Truran,  Proc.
of the Symp. on Structure and Evolution of Close
Binary Systems 73, ed. by P. Eggleton et al. (Reidel,
Dordrecht, 1976), p. 155.

18. S. Woosley and T. Weaver, Ann. Rev. Astron. Astrophys.
24, 205 (1986).

19. O. Yaron, D. Prialnik, M. Shara, and A. Kovetz,
Astrophys. J. 623, 398 (2005).

\end{document}